# Low Temperature Halide Assisted HVPE Growth of Single Crystal AlN Films

**K. Udwary[1], M. Smeaton[2], J.S. Mangum[2], R. Gannon[2], G. Dodson[1], , B. Tellekamp[2], K.L. Schulte[2] J.H. Leach[1], and J. Simon[2]**
*1. Kyma Technologies Inc, Raleigh, NC 27617*
*2. National Laboratory of the Rockies, Golden, CO 80401*
*Email : leach@kymatech.com*

**Abstract**
Low defect, single polarity aluminum nitride layers grown by halide vapor phase epitaxy (HVPE) methods have classically needed growth temperatures well exceeding 1100°C with many of the best results being grown in the range of 1400°C. These high temperatures have typically been required to obtain Al-polar AlN films with smooth morphologies and high crystalline quality. However, these elevated temperatures make the use of quartz reactor chambers difficult, as quartz begins to soften at 1200°C, and requires specially designed reactors. This study examines the use of anhydrous hydrogen chloride (HCl) gas, injected directly to the growth surface, to suppress inverted polarity grains that form at low growth temperatures. This method allows for high quality, single crystal AlN layers with a growth temperature of just 1000°C in a quartz reactor chamber. With this method, AlN layers of 5 µm thickness were grown on c-plane (0 0 0 1) sapphire, with a growth rate of 20 µm/hour, and a resulting full-width at half-maximum (FWHM) of symmetric (0 0 0 2) rocking curve reflection under 400 arcsec and asymmetric (1 0 -1 2) reflection under 1000 arcsec. Electron backscatter diffraction measurements showed the suppression of inversion domains with increasing free HCl, resulting in smoother and higher quality films.

## 1. Introduction

Aluminum Nitride (AlN) is a wide band gap semiconductor that serves a critical role in the production of high-quality ultraviolet light emitting diodes (UV LEDs), as well as high power electronics. This is due to a close lattice match between AlN and high aluminum content aluminum gallium nitride (AlGaN), and high thermal conductivity. From a crystalline standpoint, free-standing bulk AlN wafers produced from sublimation growth or physical vapor transport (PVT) methods can be very high-quality, with dislocation densities below $1x10^4$ $cm^{-2}$ [1] [2] [3]. The costs associated with such a high temperature process, as well as the lack in availability of large area substrates have prevented bulk AlN from being the preferred seed for low-cost UV LEDs and power electronics.

Thin films of AlN on sapphire (AlN templates) have shown promise as a base for UV LED production. Typically, a AlN buffer layer is grown in a high temperature metal-organic chemical vapor deposition (HT-MOCVD) process. The HT-MOCVD AlN step tends to be a time consuming and expensive process, for which several alternative growth methods have been investigated [4] [5]. Sputtered-Annealed AlN layers, using a high temperature anneal to improve the crystalline quality of the RF sputtered layers, can be a lower cost alternative, however a high impurity concentration can limit the optical and crystalline quality of such layers. Halide vapor phase epitaxy (HVPE) offers several potential advantages for AlN buffer growth over other methods, such as deposition rates more than 100 µm/h [17], large area deposition uniformity and low impurity concentrations. Carbon atoms incorporated during MOCVD growth adversely affect the transparency of AlN bulk wafers as well as thin films [6] [7], and limit the minimum free carrier concentration in nitride films necessary for high power applications[8]. HVPE growth reactors can be completely free of carbon, with growth chambers made from quartz, and the lack of the carbon containing gasses, such as the precursors used in MOCVD processes.

Typically, HVPE growth of AlN films has been limited to high temperature processes, with seed temperatures in excess of 1200°C due to low growth rates, rough morphologies and poor crystalline quality at lower growth temperatures [4] [9] [10]. Quartz components, commonly used in HVPE systems, begin to soften at 1200°C, meaning high temperature HVPE requires specialized heating of the substrate separately from typical hot wall heating, greatly complicating reactor designs, and possibly introducing carbon or other impurities into the system if an internal substrate heater were employed. Another source of impurities in a quartz-based HVPE system is the possible etching of quartz through the aluminum precursor gas. In HVPE growth, hydrogen chloride (HCl) gas is flowed over high purity aluminum metal to create an aluminum chloride precursor gas, which is then bought to the substrate surface and reacted with ammonia ($NH_3$) gas to form AlN. At temperatures above 800°C, the dominant species created by this HCl + Al interaction is aluminum monochloride (HCl + AL → AlCl) [11], which has been shown to etch quartz. It is preferable to use aluminum trichloride ($AlCl_3$) as a precursor gas, which is much more compatible with quartz at high temperatures [11]. This makes placement of an aluminum metal source internal to the HVPE growth reactor difficult due to a necessary massive temperature gradient, since the ideal temperature for the formation of $AlCl_3$ is around 400°C [12], and the substrate surface may be at a temperature above 1200°C. In this paper, an excess halide-assisted HVPE growth process is described that can produce high growth rate, high-quality AlN layers using an $AlCl_3$ precursor at a growth temperature of 1000°C through the introduction of unreacted HCl gas directly to the growth surface to promote single polarity AlN growth.

## 2. Experimental Design

The HVPE reactor used in this paper implements a vertically oriented, quartz walled growth zone capable of holding substrates up to 150 mm in diameter (Figure 1). All heating was provided from a hot wall vertical furnace, capable of temperatures up to 1100°C. A quartz injector carries the precursor gases to the wafer surface, along with carrier gasses and a channel for HCl gas. The substrate is held on a quartz platter which is raised and lowered from the growth zone with a quartz rod, which is also capable of rotation to provide improved radial uniformity. The growth chamber can operate under light vacuum, to a minimum growth pressure of about 13 kPa. For $AlCl_3$ generation, a small box furnace is located external to the growth chamber, with a sealed quartz boat located inside, which is used to hold aluminum metal pellets. The inlet of the quartz boat is connected to a gas line providing HCl gas and inert carrier for the creation of $AlCl_3$ when reacted with the aluminum metal. The box furnace is held at a temperature of 400°C to ensure that $AlCl_3$ is the dominant species produced. The outlet of the quartz boat is a stainless-steel line that is connected into the growth chamber through the quartz injector. This stainless-steel line is heated to a temperature of 200°C, to prevent the condensation of $AlCl_3$ in the line. The reactor is free of any graphite parts in the growth chamber, providing the best case for low carbon concentrations in grown films. The design of the growth chamber allows for growth of over 10 mm of material before parasitic growth begins to clog the exhaust space, compromising the vacuum of the chamber.

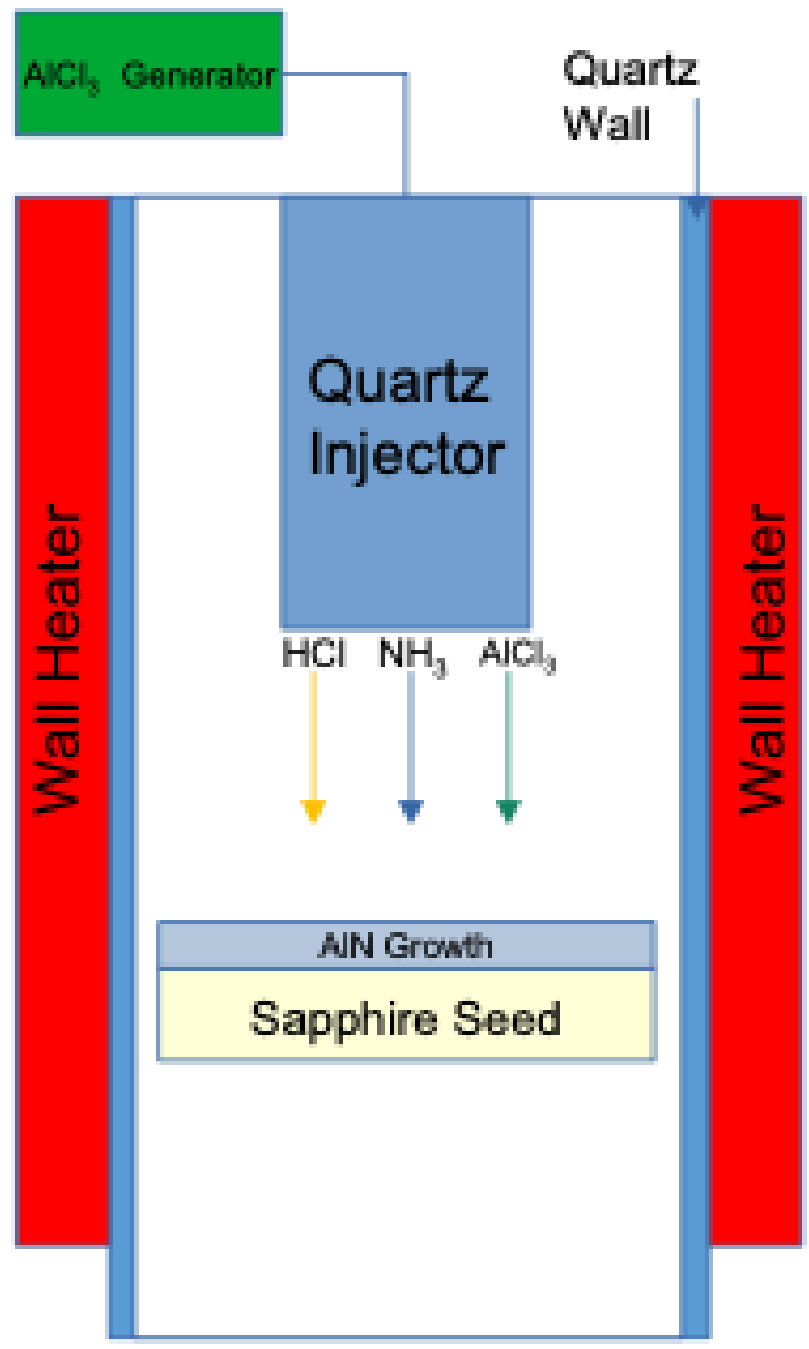

**Figure 1:** Schematic of HVPE reactor used in this study.

Samples were analyzed after growth by high resolution X-ray diffraction (XRD) using a Philips X'Pert high resolution X-ray diffractometer. Scanning electron microscopy (SEM) images were collected on each sample using a Thermo Fisher Scientific (TFS) Helios 5 CXe to assess surface and defect morphologies. Cross-sectional Electron Backscatter Diffraction (EBSD) analysis was conducted using an EDAX Clarity detector on a TFS Helios 5 Laser Hydra microscope operating at 20 kV accelerating voltage, approximately 6 nA beam current, and a pixel step size of 200 nm. Cross-sectional surfaces for EBSD were prepared using a JEOL broad beam Ar ion polishing system (IB-19550CCP). Spherical indexing cross-correlation analysis in OIM Analysis 9 was used to differentiate between N-polar and Al-polar regions in the EBSD maps. Specimens for Scanning Transmission Electron Microscopy (STEM) were prepared on the same Helios 5 Laser Hydra microscope using Argon Plasma Focused Ion Beam (PFIB). The sample surface was first coated in carbon to fill in the areas around the defects, followed by electron beam and ion beam deposited Pt and W layers, respectively. Electron transparent samples were prepared in inverted and 90-degree rotated geometries via the in-situ rotatable nanomanipulator needle to minimize curtaining artifacts from the defect. STEM images were collected on a TFS Spectra 200 STEM with a 200 kV accelerating voltage and a 24.2 mrad convergence semi-angle.

## 3. Growth Results

Initial growths were performed with only the precursor gasses ($AlCl_3$ and $NH_3$) and no added unreacted HCl, using 0.2-degree offcut, 50 mm diameter sapphire wafers. Under these conditions, with the growth temperature at 1000°C, only loose, polycrystalline AlN particles were deposited on the substrate surface. Unreacted HCl was then introduced to the wafer surface, along with the precursor gasses. It was found that at low levels of HCl, columnar micro-tower AlN growth on sapphire dominated (Figure 2c). Through XRD measurements it was found that the films had very high symmetric and asymmetric full width at half max (FWHM) peak values, with {002} reflections over 1500 arcsec and {102} reflections over 2000 arcsec. As the unreacted HCl flows were increased, the morphology of grown films improved, shown in top-down and tilted view SEM images in Figure 2. Increasing the ratio of unreacted HCl flow at 80x the flow of $AlCl_3$ precursor gasses, AlN surfaces were smoother and free of AlN micro-towers (Figure 2a). XRD peaks narrowed substantially (Figure 3) with increasing HCl/$AlCl_3$, with XRD FWHM values under 500 arcsec in the {002} reflection and around 1000 arcsec in the {102} reflection for films with a thickness of about 5 µm. It was found that with increased unreacted HCl flows, the growth rate would decrease, but even with a HCl to $AlCl_3$ flow ratio of 80:1, growth rates still exceeded 20 µm per hour.

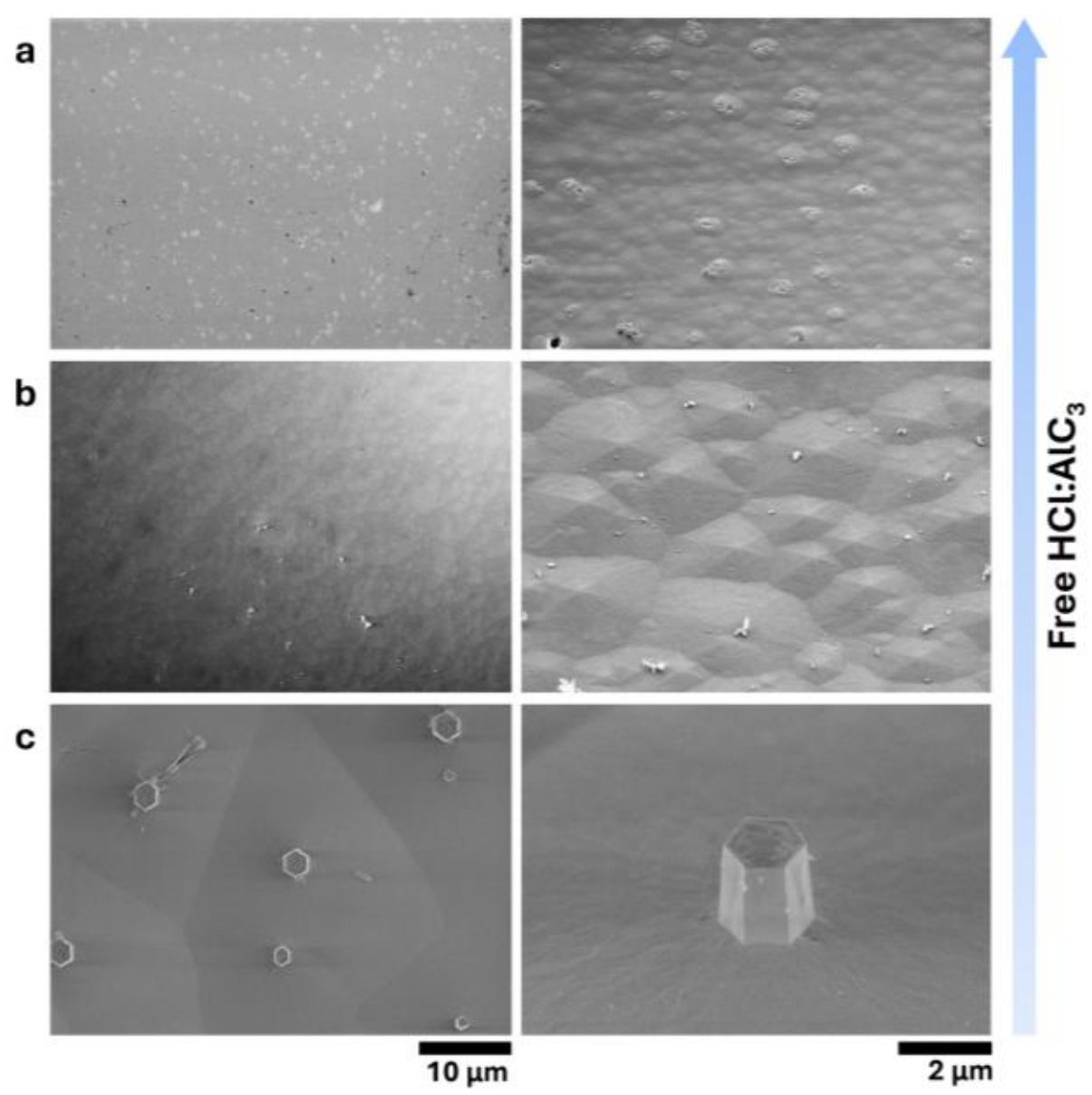

**Figure 2:** Secondary Electron (SE) SEM images of AlN films grown with varying free HCL:$AlCl_3$ ratios: a) 70.6, b) 46.2, c) 15. The left column has top-down SEM images, right-column shows a tilted SEM view of the films, where the samples were tilted to 52 degrees. SEM images in a) and b) were collected at conditions which revealed topography at 10 kV, 0.2 nA, whereas images in c) were collected at 1 kV, 13 pA.

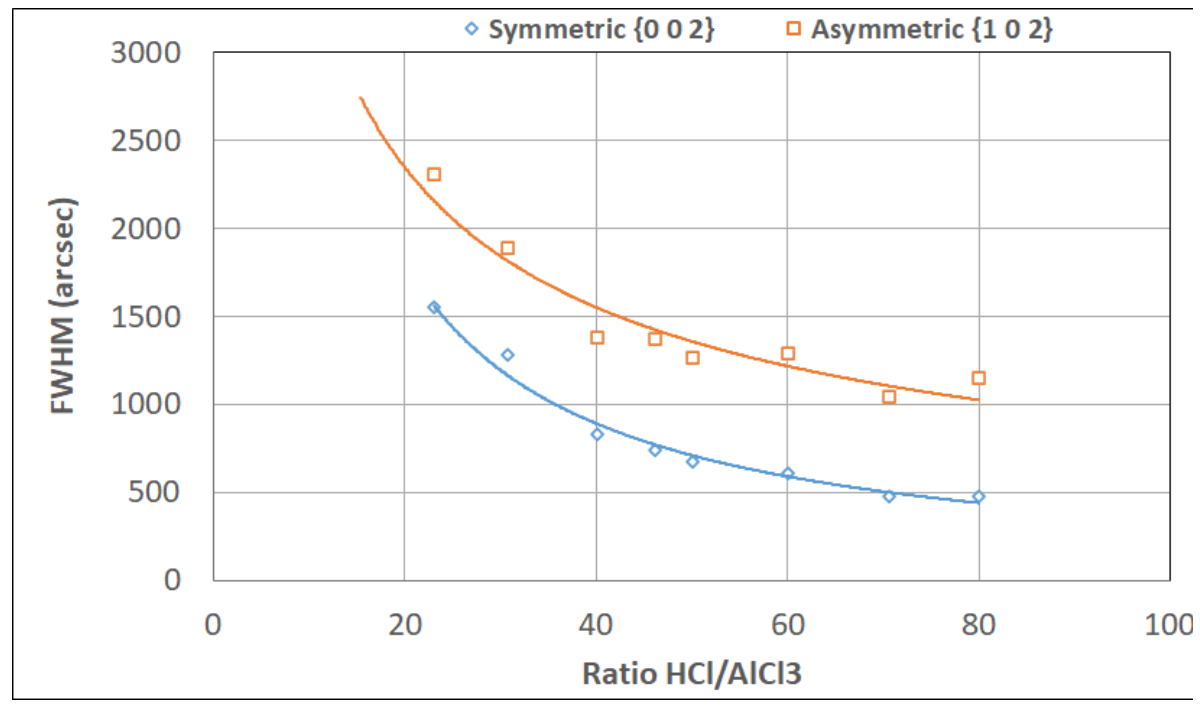

**Figure 3:** XRD diffraction FWHM for both {002} (blue) and {102} (orange) reflections.

To investigate the nature of the defects formed at low free HCl environments, EBSD measurements were performed on samples where the micro-tower defects were and were not present (Figure 4). Cross-sectional EBSD near defects shows an inversion of the polarity of the films partially through the growth. We note that differentiating between N-polar and Al-polar orientation from EBSD patterns arises from very subtle changes in intensity distribution across specific Kikuchi bands. Therefore, if a single pixel within the EBSD map spans multiple polarity domains then the resulting pattern will be a weighted average based on the fraction of each domain and the pixel will be indexed as the majority polarity within that pixel. The inverse pole figure (IPF) maps in Figure 4 indicate that the polarity of the investigated section of the AlN film is primarily N-polar but the dominant polarity inverts near the micro-tower defects, while within the interior of the defect N-poalr dominate regions do still exist. To gain a clearer understanding of the polarity at these defects, cross-sectional STEM analysis was also performed on the same sample and confirmed the presence of mixed polarity grains in the film, both within and around the micro-tower defects (Figure 5). Polarity was determined by atomic-resolution STEM imaging (Figure 5d,e) and measured for several grains in the specimen, as indicated by blue and orange markers in Figure 5c. The mixed polarity grains were observed even at the initial growth interface between the AlN film and the sapphire seed (Figure 5b), along with nanoscale defects and voids.

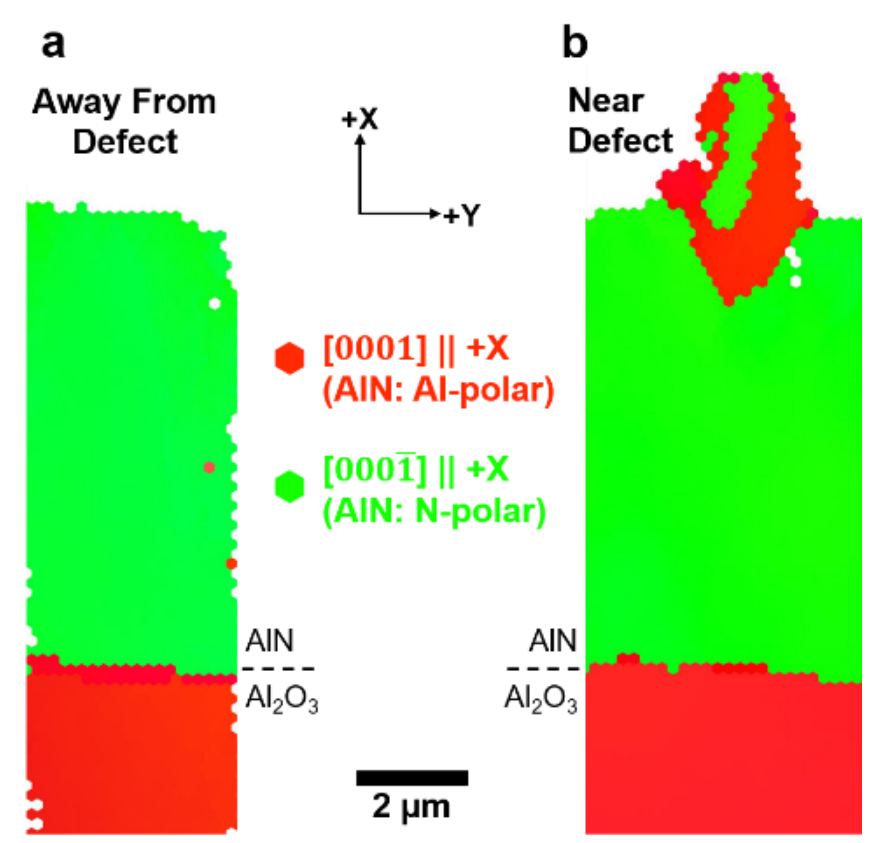

**Figure 4:** Cross-sectional EBSD IPF maps of AlN films (a) away from and (b) near growth defects present during low HCl flow rates. Pixel step size is 200 nm.

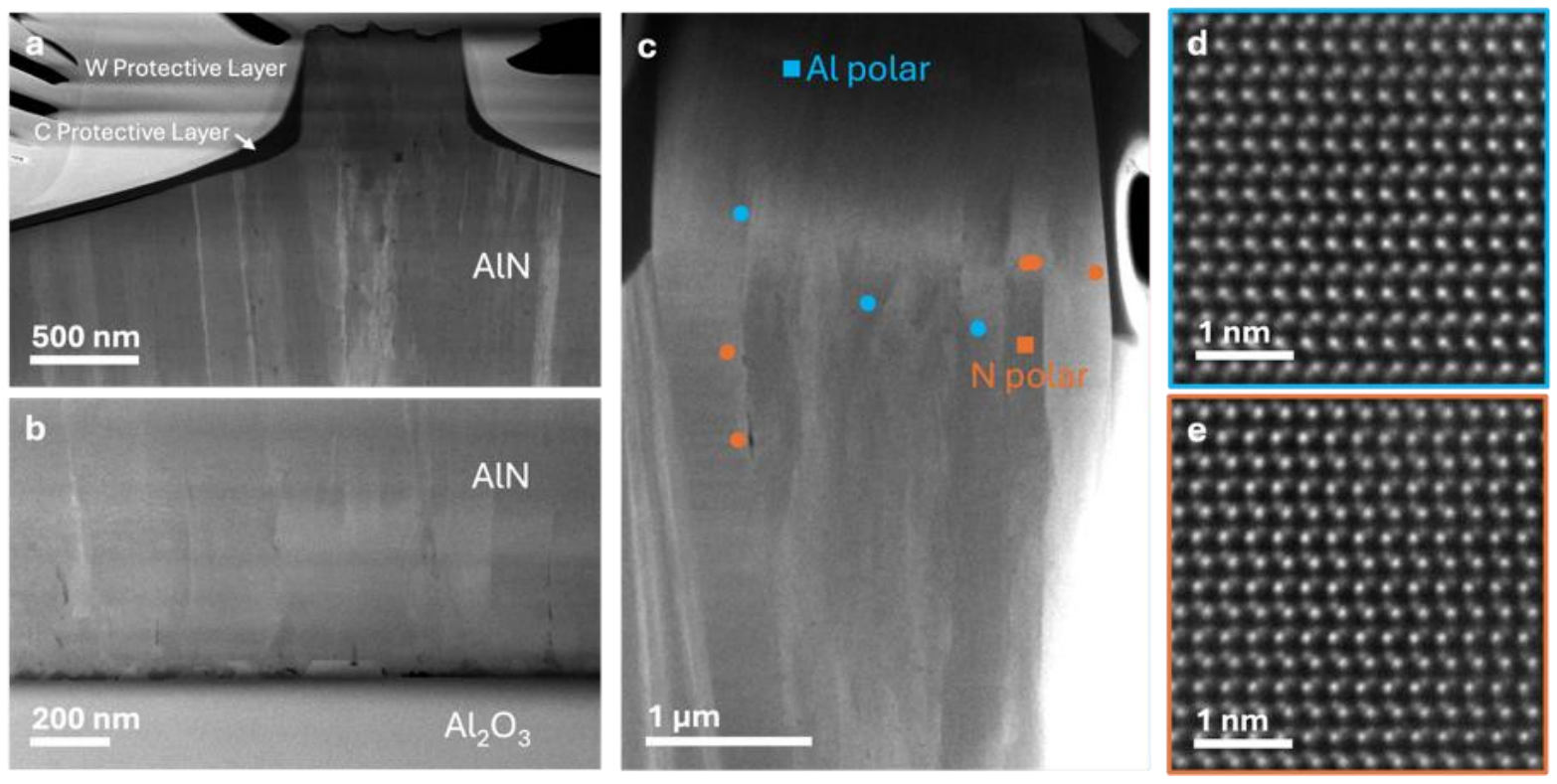

**Figure 5:** Cross-sectional STEM images of (a) an AlN micro-tower defect and (b) the AlN/$Al_2O_3$ interface. (c) STEM images of a second AlN column with verified Al and N polar regions indicated by blue and orange markers, respectively. (d,e) Atomic-resolution STEM images exhibiting Al and N polar structures, respectively, collected from the locations marked by squares in (c).

To further confirm the presence of mixed polarity grains, three films grown under differing HCl to $AlCl_3$ ratios (Table 1) were etched in an 11.7 Molar KOH solution, at 50˚C for 10 minutes. Wet etching of AlN by KOH has been shown to preferentially etch N-polar material nearly 900 times faster than Al-polar material [13]. As the HCl to $AlCl_3$ ratio is increased, AlN films showed improved crystalline quality, as determined by XRD full-width half maximum measurements on [002] and [102] reflections. This is also reflected in the KOH tests, as the AlN layers showed smaller and fewer N-polar grains as the HCl to $AlCl_3$ ratio was increased (Figure 6).

| | FWHM (ArcSec) | | | | | | | |
|---|---|---|---|---|---|---|---|---|
| Wafer | [002] | [102] | Thickness (um) | Growth Rate (um/h) | Free HCl (sccm) | Al Source HCl (sccm) | Temperature (°C) | Free HCl/$AlCl_3$ |
| AJ4367 | >1500 | >1500 | 9.72 | 58.3 | 1000 | 100 | 1000 | 30 |
| AJ5682 | 741 | 1377 | 3.5 | 30.0 | 1550 | 100 | 1000 | 51 |
| AJ5699 | 220 | 1303 | 0.36 | 21.6 | 2000 | 85 | 1000 | 70.6 |

**Table 1:** AlN on sapphire layers grown by low temperature HVPE used for KOH etching study.

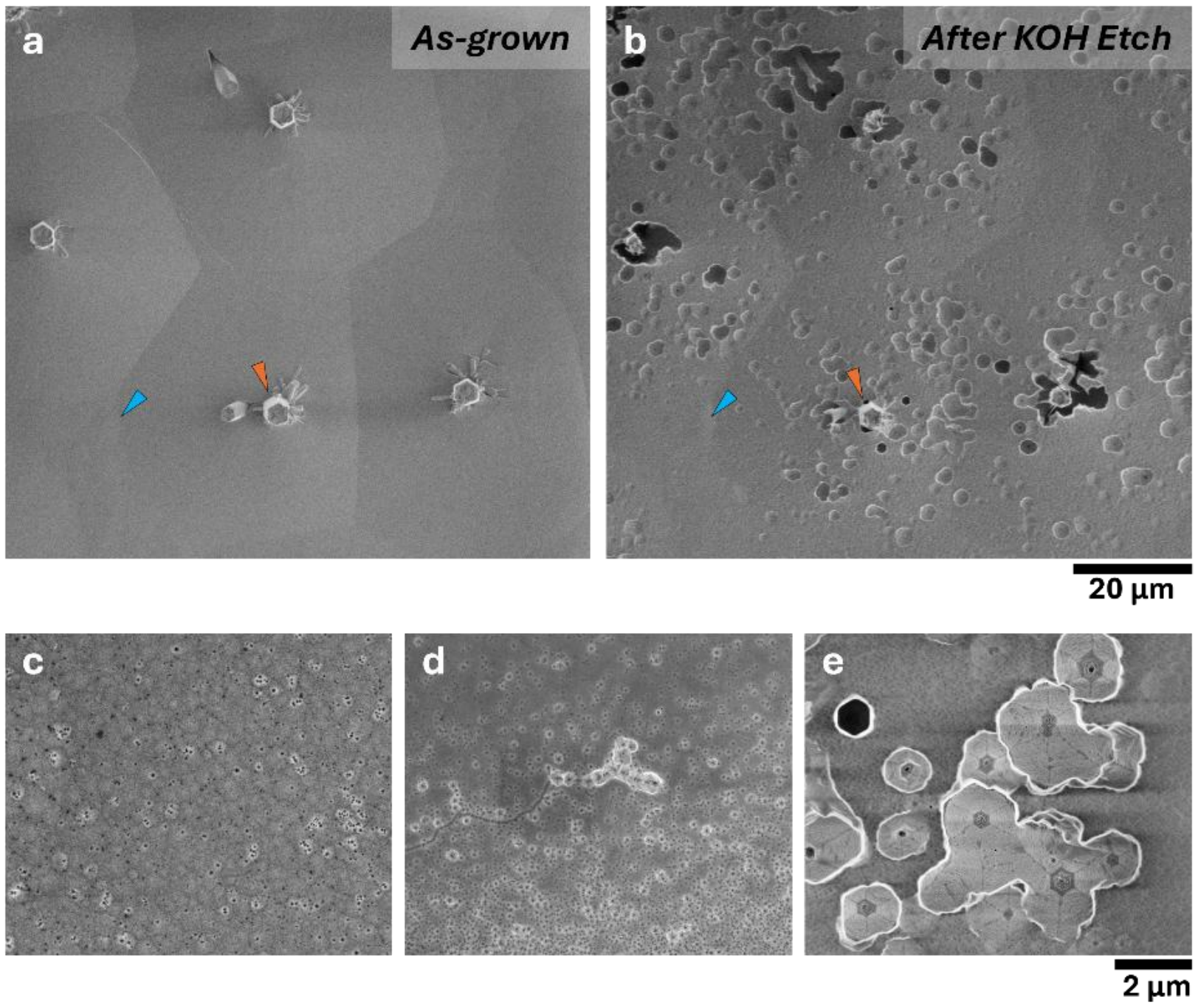

**Figure 6:** SE-SEM images of an AlN film grown with a free HCL:$AlCl_3$ ratio of 15 before (a) and after (b) KOH etching. Orange marked region contains the same defect before and after etching, where the defect area shows a distribution of etched pits which indicates mixed regions of Al-polar and N-polar AlN. Blue marked region contains the same area away from defects, which shows minimal etching indicative of a primarily Al-polar region. Images were collected at 1 kV, 13 pA with an ETD detector. Figure 6(c)-(e) contains SE-SEM images of AlN films grown with varying free HCL:$AlCl_3$ ratios: c) 70.6, d) 46.2, e) 15 after KOH etching. SEM images in c) and d) were collected at conditions which revealed topography at 10 kV, 0.2 nA with an ETD detector, whereas images in e) were collected at 1 kV, 6.3 pA with a TLD detector.

## 4. Summary

HVPE AlN layers were grown on sapphire in an all-quartz growth zone at 1000°C with the aid of unreacted HCl. It was shown that as the ratio of HCl to $AlCl_3$ precursor flow was increased, nitrogen polar grains in these films were reduced, and crystalline quality of AlN layers were increased as determined by X-ray diffraction. Many of these N-polar grains at low HCl flows seem to propagate from at the interface of the sapphire and AlN layer. It has been shown that for AlN layers grown by HVPE on sapphire the polarity of AlN grains is greatly influenced by the surface termination of the sapphire interface, and the existence of an Al(N,O) interlayer between AlN and sapphire can cause the formation of AlN with mixed polarity near the interface [14]. It has been reported that the decomposition rate of Al-polar AlN layers is lower than N-polar AlN in a hydrogen atmosphere and temperatures above 1100 °C [15], and that this allows for Al-polar domains to overgrow N-polar domains when grown at temperatures above 1100 C [16]. For the growth of AlN layers in this work, the high HCl atmosphere of growth likely has a similar effect, in that it allows for the less reactive Al-polar domains to overgrow the N-polar domains due to the high polar selective etch rate of the HCl gas, which removes the requirement of high temperatures. This lower temperature growth method allows for simpler HVPE reactor designs and potential for lower impurity concentrations in AlN films.

**Conflict of Interest:** This material is partially based upon work supported by the U.S. Department of Energy, Office of Energy Efficiency and Renewable Energy under Award Number DE-SC0013730. The material characterization and analysis work supported as part of A Center for Power Electronics Materials and Manufacturing Exploration (APEX), an Energy Frontier Research Center funded by the U.S. Department of Energy, Office of Science. This work was authored in part by NLR for the U.S. Department of Energy (DOE), operated under Contract No. DE-AC36-08GO28308. The views expressed in the article do not necessarily represent the views of the DOE or the U.S. Government.